\documentclass{optica-article}

\journal{opticajournal} 

\articletype{Research Article}

\usepackage{todonotes}
\usepackage{lineno}

\usepackage{acronym}		
\usepackage{glossaries}		

\usepackage{booktabs}

\newacronym{USyd}{USyd}{the University of Sydney}
\newacronym{ANU}{ANU}{Australia National University}
\newacronym{ESO}{ESO}{European Southern Observatory}

\newacronym{NSF}{NSF}{National Science Foundation}
\newacronym{LIEF}{LIEF}{Linkage Infrastructure, Equipment and Facilities}

\newacronym{VLTI}{VLTI}{Very Large Telescope Interferometer}
\newacronym{ATs}{ATs}{auxiliary telescopes}
\newacronym{UTs}{UTs}{unit telescopes}

\newacronym{STS}{STS}{six telescope simulator}
\newacronym{GPAO}{GPAO}{GRAVITY+ adaptive optics}

\newacronym{WFS}{WFS}{wavefront sensor}
\newacronym{ADC}{ADC}{atmospheric dispersion corrector}
\newacronym{LDC}{LDC}{longitudinal dispersion corrector}
\newacronym{AO}{AO}{adaptive optics}
\newacronym{SNR}{SNR}{signal to noise ratio}

\newacronym{OAP}{OAP}{off-axis paraboloid}
\newacronym{DM}{DM}{deformable mirror}
\newacronym{MEMS}{MEMS}{micro-electromechanical system}
\newacronym{IR}{IR}{infrared}

\newacronym{CAD}{CAD}{computer aided design}

\begin{document}

\title{Heimdallr, Baldr and Solarstein: designing the next generation of VLTI instruments in the Asgard suite}

\author{Adam K. Taras,\authormark{1,*} J. Gordon Robertson\authormark{1},
Fatme Allouche\authormark{2}, 
Benjamin Courtney-Barrer\authormark{3,4}, Josh Carter\authormark{3},
Fred Crous\authormark{1}, 
Nick Cvetojevic \authormark{2},
Michael Ireland\authormark{3}, 
Stephane Lagarde\authormark{2},
Frantz Martinache\authormark{2},
Grace McGinness\authormark{3}
Mamadou N'Diaye\authormark{2},
Sylvie Robbe-Dubois \authormark{2},
Peter Tuthill\authormark{1}
}

\address{\authormark{1}Astralis-Usyd, Sydney Institute for Astronomy, School of Physics, University of Sydney, NSW 2006, Australia\\
\authormark{2}{Université Côte d’Azur, Observatoire de la Côte d’Azur, CNRS, Laboratoire Lagrange UMR 7293, 28 Avenue Valrose, F-06108 Nice Cedex 2, France}\\
\authormark{3}{Research School of Astronomy and Astrophysics, Australian National University, Canberra 2611, Australia}\\
\authormark{4}{European Southern Observatory, Alonso de Córdova 3107 Vitacura, 19001, Santiago, Chile}}

\email{\authormark{*}adam.taras@sydney.edu.au} 


\begin{abstract*} 

High angular resolution imaging is an increasingly important capability in contemporary astrophysics. Of particular relevance to emerging fields such as the characterisation of exoplanetary systems, imaging at the required spatial scales and contrast levels results in forbidding challenges in the correction of atmospheric phase errors, which in turn drives demanding requirements for precise wavefront sensing. Asgard is the next-generation instrument suite at the European Southern Observatory's Very Large Telescope Interferometer (VLTI), targeting advances in sensitivity, spectral resolution and nulling interferometry. In this paper, we describe the requirements and designs of three core modules: Heimdallr, a beam combiner for fringe tracking, low order wavefront correction and visibility science; Baldr, a Zernike wavefront sensor to correct high order atmospheric aberrations; and Solarstein, an alignment and calibration unit. In addition, we draw generalisable insights for designing such system and discuss integration plans. 
\end{abstract*}

\section{Introduction}
The Very Large Telescope Interferometer (VLTI) operated by the European Southern Observatory (ESO) at Mt Paranal in Chile is a major facility dedicated to study of celestial objects at the highest angular resolutions \cite{Kervella2000Vinci, petrov2007AMBER,GRAVITY2018}. We describe here three of the central modules within the Asgard suite of instruments \cite{Martinod_JATIS} designed to advance the VLTI's capability in sensitivity, spectral resolution, and nulling interferometry (to cancel starlight when searching for companions at high contrast).

\begin{figure}[htbp]
\centering\includegraphics[width=0.40\textwidth]{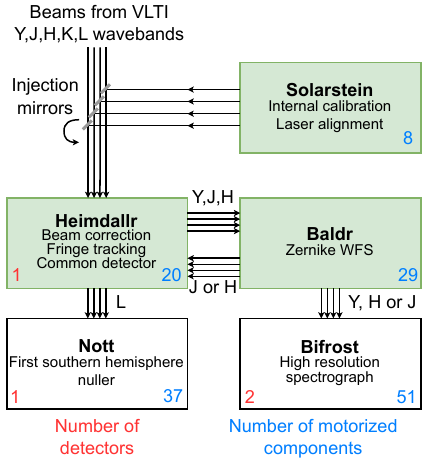}
\caption{Block diagram of the Asgard instrument suite. 
Modules covered in this paper are shaded green. 
Heimdallr contains common optics that correct the beam aberrations, sensing K band while splitting and distributing the other wavebands. The Y, J and H bands go to Baldr on the upper level, which
can operate in either J or H, with the Bifrost spectrograph using H or J, and Y. Baldr is on the upper level, but it shares a common detector with Heimdallr and hence the beams return. Calibration and alignment beams are be generated by Solarstein, with injection mirrors used to switch between sky and the internal source. }
\label{fig:overview_block}
\end{figure}

\begin{figure*}[htbp]
\centering\includegraphics[width=0.88\textwidth]{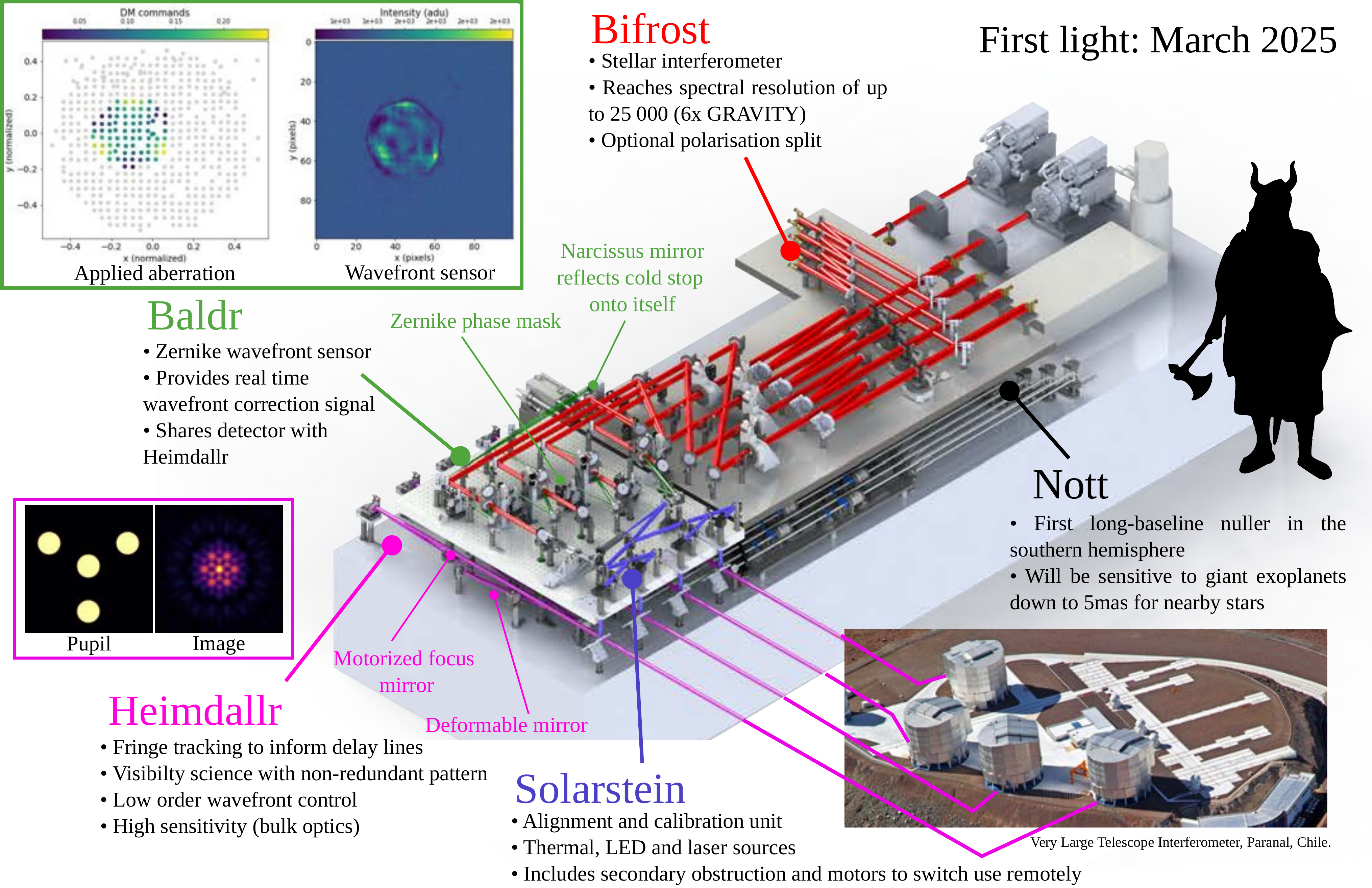}
\caption{Summary of the Asgard suite and the key functions of each module. Black silhouette for scale. Baldr inset (green) shows a preliminary mask being used to sense the wavefront (right) from a beam with a known aberration (left). Heimdallr inset illustrates the non-redundant pupil pattern and the corresponding image on the detector for a single wavelength. }
\label{fig:summary_cad}
\end{figure*}

The key contributions of this work are:
\begin{itemize}
    \item We detail the requirements and designs of the Heimdallr, Baldr and Solarstein modules, providing a case study in developing interferometric, adaptive optics, and calibration systems that is valuable for other instruments; and
    \item We present generalisable insights to working with large optical systems, in particular dealing with the challenges of multiple beam, sequential optics.
\end{itemize}
For each module, we present the driving requirements followed by optical and optomechanical designs. We highlight decisions that could be of wider interest for instrumentation. Finally, we present system-wide insights and integration plans.

The VLTI will provide beams from either the four 1.8m Auxiliary Telescopes (ATs) or the four 8.2 m Unit Telescopes (UTs). After adaptive optics correction and path length compensation the beams reach the Asgard optical table.

\autoref{fig:overview_block} depicts a block diagram of the Asgard suite, with each module accepting 4 beams, one from each telescope. The large number of detectors and motorised components reflect the complexity and capability of the system, with a plethora of observing modes available. Heimdallr, Baldr and Solarstein play critical roles in enabling the other Asgard modules to function: delivering well-conditioned, corrected and cophased beams for observations on sky or, in the case of Solarstein, use daytime alignment and calibration. The suite as a whole spans the range of the infrared wavelengths observable from the VLTI: Y $0.96 - 1.08\mu$m; J $1.1 - 1.33 \mu$m; H $1.5 - 1.88 \mu$m; K $1.95 - 2.35\mu$m; L $3.2 - 4 \mu$m.

\autoref{fig:summary_cad} shows a labelled render of the suite. All modules fit on an optical table (previously occupied by the AMBER instrument) in the beam combining facility, with an upper level added to accommodate the extensive optics required for the full instrument suite.

\section{Heimdallr}

\subsection{Goals and requirements}

Heimdallr is the first module the VLTI beams encounter in the Asgard suite. Hence, it conditions the beams in common optics, compressing the beams and changing their height above the table from 200mm to 125mm for stability of mounts. The common optics also features active components to apply wavefront corrections and cophasing. All wavebands for other modules must then be split and routed to their respective instrument. Finally, Heimdallr also constitutes the final destination for K band light which is recombined yielding dual-use signals for fringe tracking/cophasing and interferometric science operation. 

We distil these requirements further. In order to fringe track effectively, a non-redundant pupil image is formed on the detector in two wavebands yielding cophasing information robust against phase wrapping. High cadence fringe tracking delivering corrections at frequencies $\ge 2$ kHz, and offloading larger piston errors to the VLTI delay lines at $30-100$ Hz is required. This is enabled through the use of a 4-way beam combination scheme configured as a non-redundant pupil, where unique parts of the power spectrum map one to one with relative phases between beams \cite{Ireland2018Image}. Furthermore at the longer wavebands the instrument must effectively mitigate the rapidly rising spectrum of thermal noise from the warm optics, laboratory and sky.

\subsection{Overview of the Heimdallr instrument}

Heimdallr is an image-plane beam combining interferometer that formats the four individual telescope beams into a Fizeau-type non-redundant 2D array layout (\autoref{fig:summary_cad}, `Pupil' inset) \cite{Ireland2018Image}. 
Given the detector pixel size of 24 $\mu$m, fringe sampling at or above the Nyquist limit requires the outer triangle of three beams to converge at $1.04^\circ$ relative to the central one, and each should reach the detector at $f/121.5$. The design requires at least some multi-wavelength capability in order to resolve phase ambiguities – here we use the minimum of just two wavelengths so as to optimise fringe tracking sensitivity, by splitting the K band into K1 (1.95 – 2.15 $\mu$m) and K2 (2.15 – 2.35 $\mu$m). Separate interferograms are formed from the 4 beams in each of K1 and K2 \cite{Ireland2018Image}.

\begin{figure}[htbp]
\centering\includegraphics[width=0.8\textwidth]{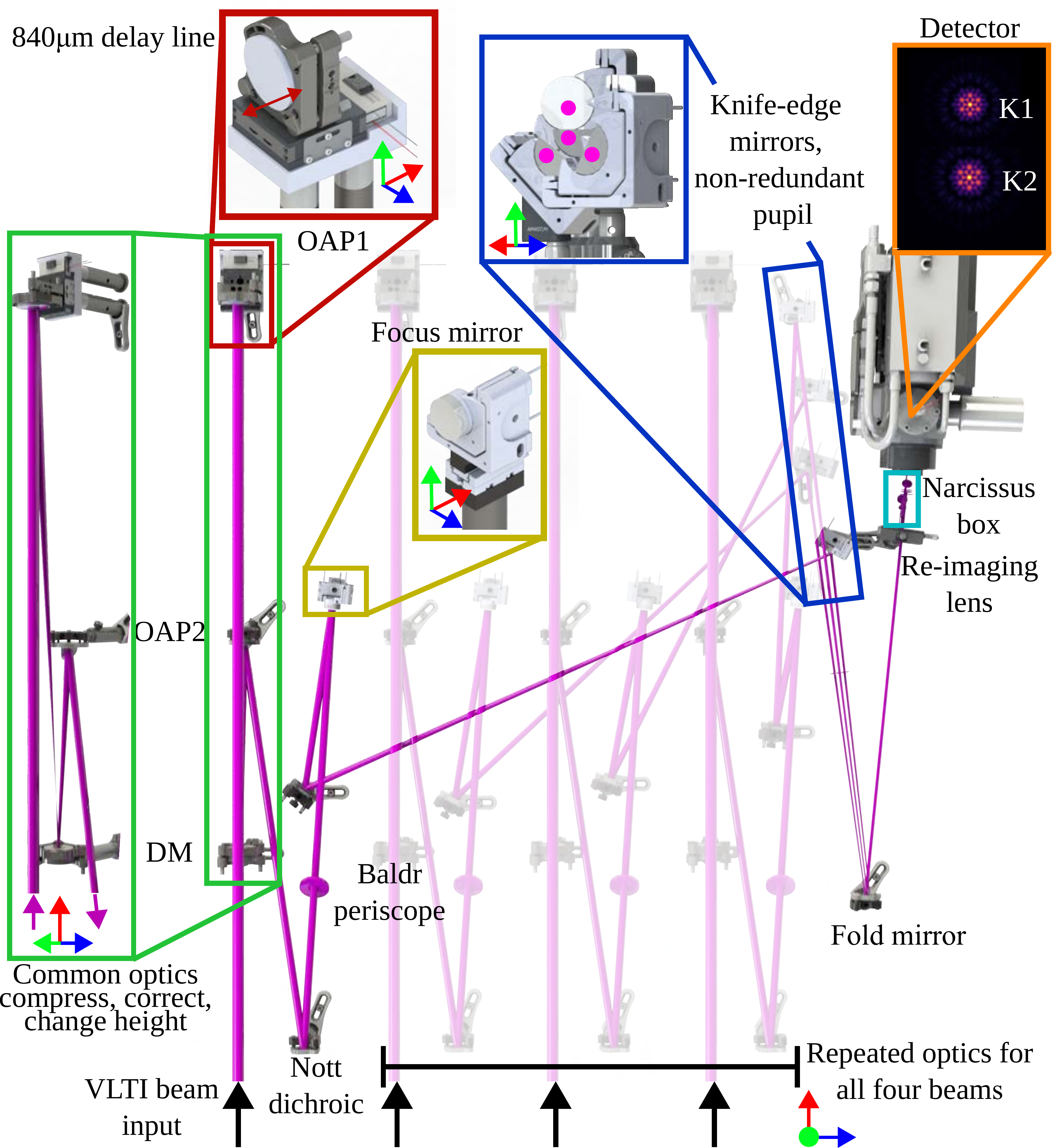}
\caption{Top view of Heimdallr optical layout, with repeated paths for each of the four beams partially transparent for clarity. The common optics (OAP1, \gls{DM}, OAP2) compress, delay, correct and change the height of the beam. After dichroics pick off NOTT and Baldr beams, the motorized focus and knife edge mirrors align the beams in a non-redundant mask pattern (blue box) that, when imaged on the detector, enables low order wavefront control and visibility science. Fold mirror labels are omitted for clarity. Details of the Narcissus box are shown in \autoref{fig:Narc_box}.}
\label{fig:heimdallr}
\end{figure}

The four 18 mm diameter AO-corrected and collimated VLTI beams enter Heimdallr, with the pupil 2510 mm beyond the edge of the optical table (see \autoref{fig:heimdallr}). 
Prior to splitting wavebands out to different modules, starlight spanning the full IR band first encounters the ``common optics'' (replicated 4 times for input beams).
Two off-axis paraboloids (OAP1 and 2) reduce the beam diameter to 12 mm, as used in the remainder of Asgard. Notes regarding the use of OAPs in the Zemax optical design system are given in the supplementary material.

The focal length of OAP 1 is chosen so that it re-images the pupil on to the \gls{DM} surface at the chosen diameter of 3.8 mm. The Deformable Mirror needs to operate at an update rate of at least 2\,kHz (i.e. less than 200\,$\mu$s latency), in order to act as both a fast piston and adaptive optics actuator for at least 50 modes and have a piston stroke of at least 2\,$\mu$m. The standard Multi-3.5-DM from Boston micromachines meets these requirements. The beam leaves OAP2 at 125 mm above the table and then encounters the dichroic to NOTT (marking the end of the ``common optics''), which transmits the L band light to suitably-placed fold mirrors (not shown).

The next component is a $45^\circ$ inclined dichroic, which sends the YJH light up a periscope to the upper level breadboard, for Baldr and Bifrost, while transmitting the K beam to the rest of Heimdallr. 
After bypassing the Baldr periscope via the dichroic, each telescope’s K beam encounters a spherical mirror of long focal length (2.0 m). To minimise aberrations, the angles of incidence are as small as possible -- the largest is $3.498^\circ$. 
As the beams gradually converge, they go via a steering mirror and then to an array of knife-edge mirrors which form the 4 beams into the convergent triad around the central fiducial beam. The mirror placements preserve path length equality.

\begin{figure*}[htbp]
\centering\includegraphics[width=0.99\textwidth]{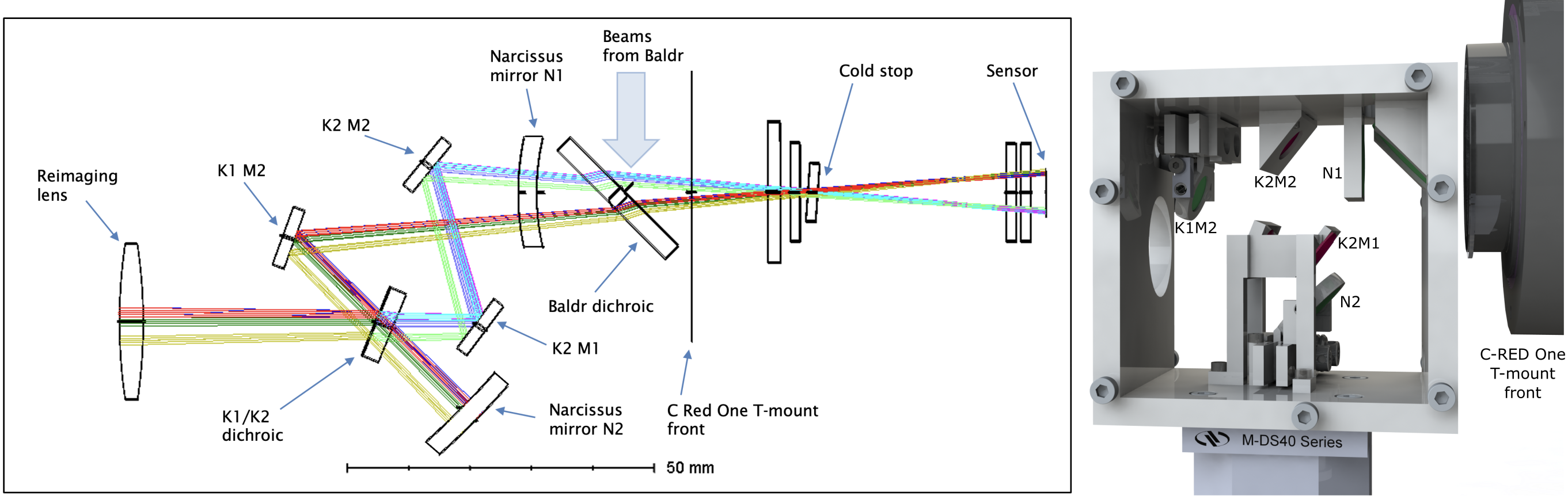}
\caption{(left) Side view of the final elements before the detector, including the filters within the C-RED One. The assembly immediately in front of the C-RED One is referred to as the ‘Narcissus box’. The N1 optic has holes drilled to allow starlight to pass through. Note that the rays shown reaching N2 are notional -- only thermal radiation to and from the cold stop area should exist in this space. (right) Side view of the mechanical design of the Narcissus box, mounted on a XYZ translation stage. The model includes custom mounts for tip/tilt adjustment of N2 and K1M2. }
\label{fig:Narc_box}
\end{figure*}

The C-RED One detector is internally cooled, but none of the external optics are cooled. To minimise the thermal background reaching the detector, a number of steps have been taken:
\begin{enumerate}
\item	The C-RED One internal cold stop has diameter 2.145 mm;
\item	Two Narcissus mirrors will be used, to reflect the cold stop back on itself (\autoref{fig:Narc_box}). Each Narcissus mirror will have a radius of curvature equal to its optical distance from the cold stop. Mirror N1 has 8 holes drilled in it to allow the beams through (\autoref{fig:thermal}); and
\item	The entire C-RED One is tilted by $3.197^\circ$ relative to the incoming beam, so that the Heimdallr spots are received towards the edge of the sensor area. This restricts stray radiation coming through the holes in N1 to one side of the sensor, leaving the other side uncontaminated for Baldr.
\end{enumerate}

A fold mirror directs the beams to a reimaging lens, which forms a pupil image close to the Narcissus mirror N1 (so the holes in N1 can have the minimum size) while at the same time reimaging the beams to the correct individual $f/$\# and correct convergence angle at the sensor plane.
After the reimaging lens, a dichroic reflects K1 while transmitting K2 (\autoref{fig:Narc_box}). Fold mirrors then direct the beams through the narrow cold stop and on to the desired area of the sensor plane, while keeping the two sets parfocal. 
A further dichroic just outside the C-RED One housing receives the beams coming down another periscope from the Baldr breadboard, and directs them through the cold stop and on to the opposite side of the sensor, away from the Heimdallr area.
The Heimdallr optical components are listed in Table S1 in the supplementary material.

\subsection{Field of view}
The required field of view varies substantially between various parts of the system. To allow Bifrost’s off-axis mode, the desired Field of View (FoV) is $\pm 1''$ at the Unit Telescopes which translates to $\pm455''$ in the 18 mm pupils and $\pm683''$ at the 12 mm beams. It is also necessary to consider atmospheric dispersion, especially at the front-end of the system.  The full $\pm1''$ is available for NOTT at its dichroics, and on the upper level as far as the Bifrost dichroics. 

After passing the Baldr/Bifrost dichroic, the Heimdallr requirement for FoV is governed by single-object diffraction, judged to be satisfied by $\pm 3 \lambda/D = 0.177''$ (at the UTs) until the Narcissus box and $\pm 1 \lambda/D = 0.059''$ within the Narcissus Box and detector.

\subsection{Delay line function}
\label{sec:delay_line}
A requirement for successful interferometry is that path delays for the interfering beams must be equal within the coherence length, despite geometrical path differences which may be tens or hundreds of metres. The VLTI delay line precedes Asgard, and will largely compensate for the path differences in real time. However, Asgard is required to incorporate a small delay line capability, independent of the VLTI. This will enable delays which exceed the DM's range to be rapidly offloaded to the Asgard internal delay line, thus simplifying the interface with the VLTI system. A delay range of $\pm$ 0.8 mm will be sufficient for this purpose, based on the outer scale of turbulence measured at Paranal \cite{2010A&A...524A..73D}.

Ideally a delay line should reflect the beam back exactly parallel to the incident beam, so there is no lateral displacement (pupil shift) as the delay changes. It should also be in a collimated beam so that there are no effects on focus of the image. However, there is no such ideal optic naturally within Heimdallr, and we wish to avoid losses incurred by adding further surfaces. The most suitable optic for actuation is OAP1, for which even a large delay of 1 mm (surface shifted by 0.5 mm) causes a modest displacement on the detector of $\sim100 \mu$m.

The spherical focus mirrors will also be movable over a range of a few mm. This will enable a differential delay to be inserted between Heimdallr and Bifrost/NOTT, for the purpose of static cophasing the three instruments. Moving the focus mirror by up to $\pm$ 2 mm (hence $\pm$ 4 mm differential delay) causes minimal degradation of the image quality.

\subsection{Thermal Background}

\begin{figure}[htbp]
\centering\includegraphics[width=0.85\textwidth]{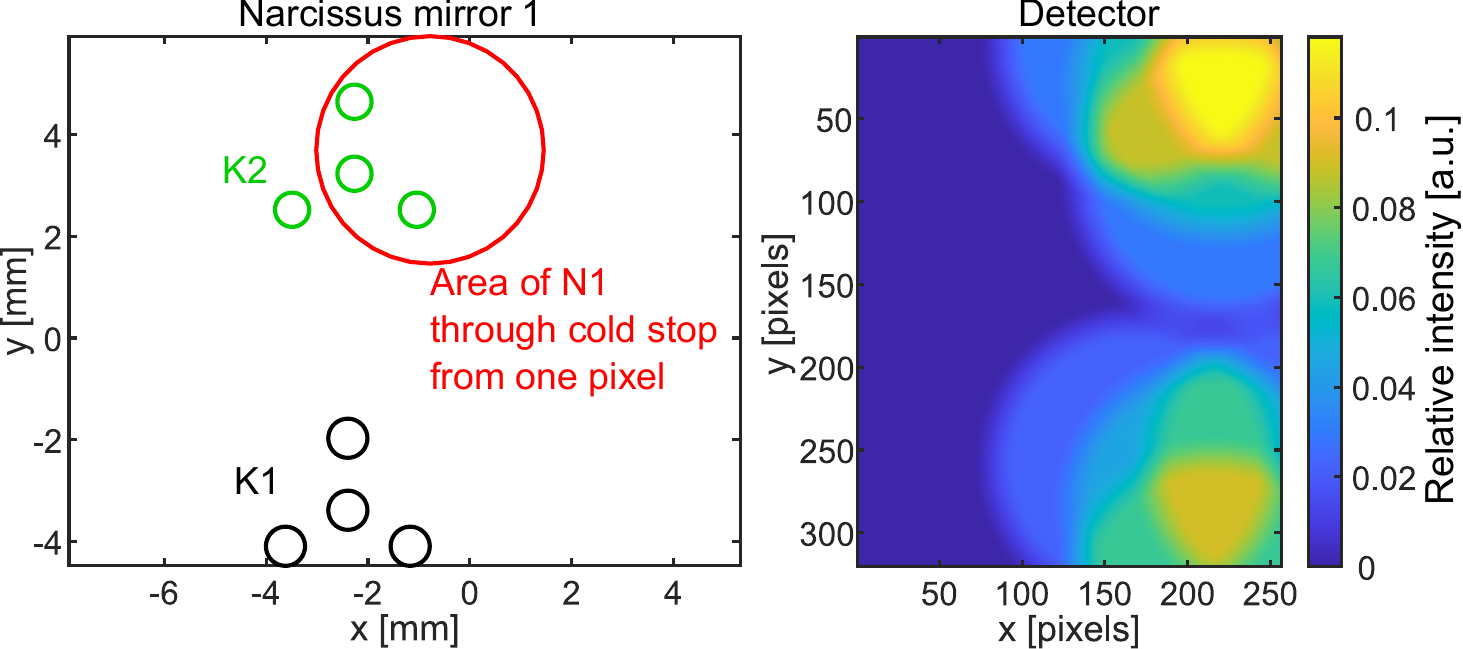}
\caption{Narcissus mirror design. (left) Map showing part of N1, with K1 holes in black, K2 in green. The red circle shows the area of N1 seen through the cold stop by one particular pixel on the sensor. (right) The colour map shows the relative intensity of thermal radiation reaching the sensor ({\it i.e.} solid angle of overlap with the holes in N1). K1 is at the top, K2 at the bottom (opposite to the left-hand panel, since the beams cross at the cold stop). Patterns are not centred to allow room for Baldr measurements on the same detector.}
\label{fig:thermal}
\end{figure}

The C-RED One sensor receives background thermal radiation through the 8 holes in the Narcissus mirror N1 (4 for each of K1 and K2). For the Heimdallr fringes this is unavoidable, but for placement of the Baldr spots we seek a place on the detector with minimal background. Each pixel on the sensor sees just a part of N1 through the cold stop, and the amount of thermal radiation reaching that pixel is proportional to the total solid angle of holes in N1 that lie within the pixel’s footprint on N1.  A simulation was used to find the overlap area as a function of position on the sensor. The resulting distribution (right hand panel of \autoref{fig:thermal}) shows that there is indeed a substantial area to the left side of the sensor, which cannot see any of the holes in N1 and hence is expected to have lower thermal background.

\subsection{Heimdallr image quality}
The quality of delivered images is important because any degradation can reduce the fringe visibility. A check of the Optical Path Difference at the image plane shows that the maximum error is only $0.012 \lambda$  ({\it i.e.} $\lambda/83$).  Aberrations will be increased slightly if OAP 1 is displaced from its nominal position when acting as an internal delay line (Section \ref{sec:delay_line}) but the wavefront error is still small, at a maximum of $0.03 \lambda$.

A diffraction analysis has also been carried out, and shows a clean Airy disc pattern in the image plane. At the cold stop, the wings are slightly greater but the image is still strongly centrally concentrated. Only 1-3 \% of the beam flux is vignetted there (for central or peripheral beams respectively -- see \autoref{fig:summary_cad}).

\subsection{Optomechanical design}

We now turn to consider notable parts of the optomechanical design. We created a Solidworks model of the system as a whole, using sub-assemblies for each optical component. The volumes of mounts and optics are measured with respect to beams to ensure sufficient clearance, avoiding vignetting. A detailed explanation of the methods used to integrate multiple Zemax modules in Solidworks is discussed in \autoref{sec:discussion}. The critical knife-edge configuration (blue inset in \autoref{fig:heimdallr}) is verified using this process.

We have also adapted our design to include system wide delay lines on OAP1. Initially, piezo stick-slip motors were tested, but showed undesirable tip effects and limits with actuation speed. The final design instead employs a mechanically  amplified piezoelectric actuator (PK2FVF1), with 420$\mu$m travel. This drives a sturdier translation stage as shown in the red inset in \autoref{fig:heimdallr}.

The alignment of the system through the Narcissus box requires, for effective thermal suppression, strict repeatability requirements. If beams do not land on the detector, automated adjustments of the motors upstream become difficult to debug. Hence, we have selected CONEX-AG-M100D motors for the spherical and knife-edge mirrors in Heimdallr, which meet the resolution requirement enforced to pass the beam through the Narcissus mirror holes and cold stop without vignetting. The direct-drive technology ensures the position is held even after a power cycle, ensuring beams in Heimdallr will consistently avoid being vignetted through the Narcissus box during the alignment process. 

In addition, we consider the alignment of the Narcissus box to be a critical step. We use a custom frame (\autoref{fig:Narc_box}, right) that constrains almost all optics in the correct position and orientation by design, and instead allow the box as a whole to easily be adjusted with respect to the detector using an XYZ manual translation stage. The beams are steered using motors outside the box to actuate the fold mirror and re-imaging lens, minimising the number of moving parts in this compact design.

\subsection{Summary}

The Heimdallr beam combiner fulfils the beam conditioning required for the entire Asgard suite, manages the waveband splitting and finally interferes all four beams on the detector in a non-redundant pattern, enabling visibility science as well as low order wavefront correction and fringe tracking. The design of this instrument is a showcase for the challenges but also opportunities with working in the thermal background environment of the K band and with four telescopes.

\section{Baldr}
\subsection{Goals and requirements}
Baldr is a Zernike wavefront sensor (ZWFS) that operates on the beam from each telescope independently with four optically identical sets of optics. Its wavefront data will be used to drive the deformable mirrors in each of the four beams. The operating modes of Baldr are driven by Bifrost science requirements, optimising fibre injection by sensing higher order aberrations. Most notably, the system will be able to sense the wavefront in either the J and H bands. Furthermore, Baldr deploys two different sensitivity modes -- one where around 70 modes are sensed using $12\times12$ pixels, and another where the beam is focused to only $6\times6$ pixels. The latter case increases the number of photons per pixel avoiding read-noise limitations at the cost of reducing the number of effectively controlled modes. This enables \gls{AO} corrections for fainter targets. For operation on the 8m unit telescopes, Bifrost and Baldr require a common \gls{ADC}. Finally, for compactness, synchronisation and cost, Baldr and Heimdallr share a common detector.

\subsection{Background: Zernike wavefront sensor}

A ZWFS was selected for the Baldr AO module of Asgard based on consideration of technology readiness and photon-noise sensitivity. In particular:
\begin{itemize}
    \item ZWFS have been shown to be the optimal wavefront sensor in-regards to photon noise sensitivity \cite{guyon2005aolimits};
    \item ZWFS are a proven technology - they are standard practice in microscopy (known as phase contrast methods), have been proven to work on-sky in astrophysical applications including on VLT/SPHERE with Zelda \cite{vigan2019zelda_on_sky}, and are being planned for use in the next generation telescopes such as ELT/HARMONI \cite{2022SPIEharmoni}; and
    \item Baldr will be a second-stage AO system. Therefore the inherently limited dynamic range of ZWFS's will be largely overcome, with the first stage AO correction being done by Naomi or GPAO for the AT's or UT's respectively. These routinely provide H-band Strehl ratios in the 20-90\% range which makes the input beam within dynamic capture range of a ZWFS.
\end{itemize}

The working principle of a ZWFS is to focus the input light onto a phase mask which applies a non-zero phase shift across part of the core of the point spread function. The interference between the phase-shifted (so-called reference) photons with the non-phase shifted photons in a pupil plane after the phase mask then creates the desired effect of a ZWFS to make the output pupil plane intensity proportional to the phase aberrations across the input pupil. While the lower end of this initial input Strehl ratio may be outside the linear range of the ZWFS, either a type of lucky imaging and/or non-linear phase reconstructors \cite{ndiaye_2013_zelda} could be used to initially close the AO loop and bring it into a linear operating range.

By prior knowledge of the input parameters and/or calibration on a known calibration source, a set point can be calibrated, or analytically derived to construct the control signal. 
Many previous authors have shown how ZWFS are the most photon-sensitive WFS. See, for example, Figure 10 in \cite{guyon2005aolimits} and Figure 6 in \cite{chambouleyron2021dotsize} for a sensitivity comparison between various WFS's using different sensitivity definitions. In particular, ZWFS have a theoretical sensitivity a factor of two better than pyramid WFS's and at least a factor of three better than Shack–Hartmann WFS's \cite{guyon2005aolimits}. The sensitivity to particular modes can be optimised by selectively choosing different diameters and/or depths of the phase shifting region on the phase mask, typically with tradeoffs between sensitivity of higher vs lower order modes. Therefore it can be useful to consider various phase masks (or various phase shifting dots on the same phase mask) for different observing modes.

\subsection{Phase mask sizing}
In order to compare design parameters, we use the definition of wavefront sensor sensitivity from \cite{fauvarque_2016_sensitivity_formalism_2}. 

\begin{figure}[htbp]
\centering\includegraphics[width=0.85\textwidth]{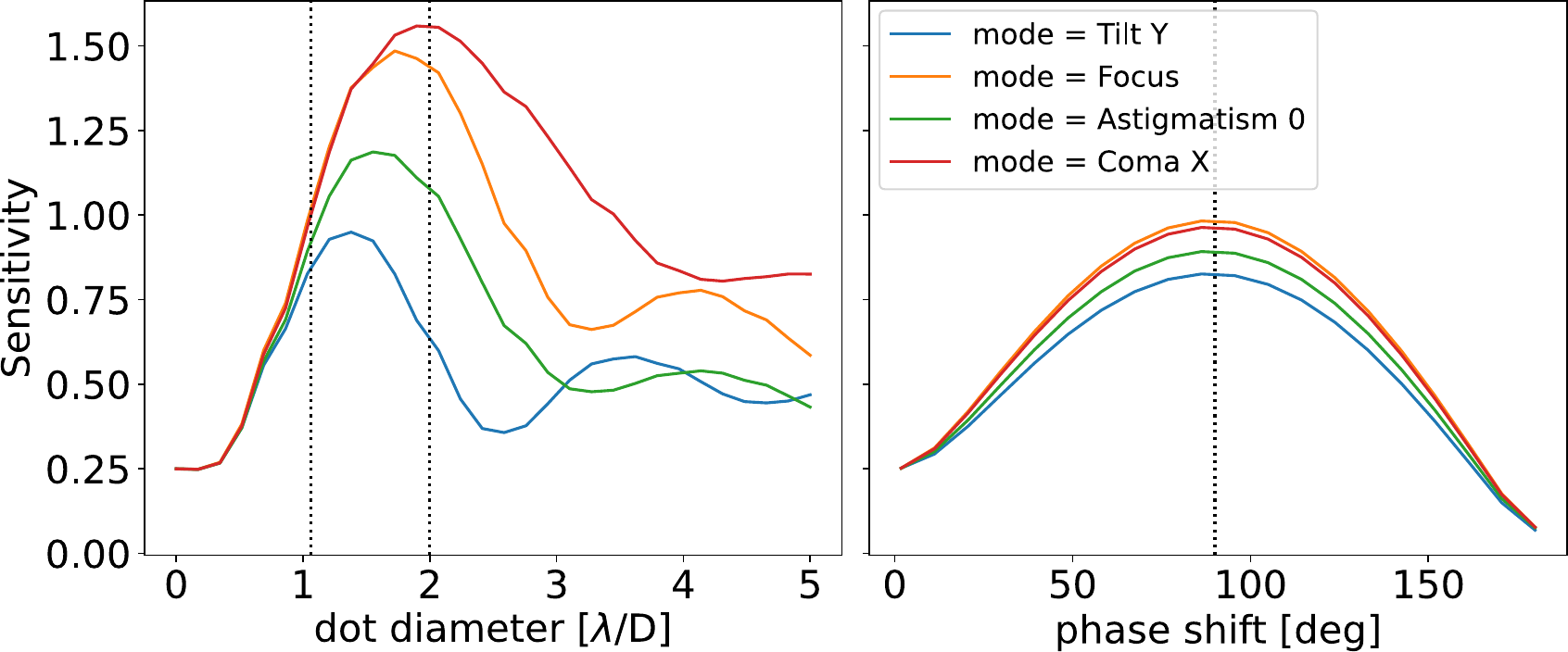}
\caption{Phase mask design tradeoff for sensitivity to Zernike modes. (left) Sensitivity as a function of phase mask dot diameter for some monochromatic Zernike modes with phase mask depths corresponding to a $\pi/2$ phase shift. The black dashed lines highlight the sensitivities for the traditional 1.06 $\lambda/D$ vs a 2 $\lambda/D$ diameter phase mask. (right) Sensitivity as a function of phase mask phase shift for some monochromatic Zernike modes with phase mask diameter equal to 1.06 $\lambda/D$. The black dashed line indicates $90 ^\circ$ where the peak sensitivity is found for all modes.}
\label{fig:sensitivity_vs_dot_diameter}
\end{figure}

Considering a monochromatic source we analyse the modal sensitivity vs the depth and diameter of the phase shifting dot on the phase mask. \autoref{fig:sensitivity_vs_dot_diameter} shows simulated sensitivities over an indicative parameter space for the monochromatic case. 
A similar simulation including chromatic effects indicated only slight linear shift in optimal phase shift at the bandpass central wavelength vs spectral bandwidth, and similar results for dot diameter sensitivity. Based on this analysis, the phase mask design shall include two columns of phase shifting dots corresponding to the H and J band observing modes respectively. The dot depths across each column will be the same, corresponding to a 90 degree phase shift at the central wavelength of the H or J bands to maximise sensitivity. Depending on manufacturing complexities each column will have various dots of different diameters between $1-2 \ \lambda/D$ for the respective central wavelength. This will allow user selection of modal sensitivities.

\subsection{Optical design}

The optics for Baldr are shown in \autoref{fig:baldr}, with components listed in Table S2 of the supplementary material.  The beams covering YJH bands are directed up a periscope by the dichroic mentioned above. The periscope contains an \gls{ADC}, which consists of a pair of zero-deviation Risley prisms.  After arrival at the upper level breadboard, the beam reaches the Bifrost dichroic, which is motorised and can reflect either H or J bands. The reflected beam is focused by an OAP. At the star focus we place the Zernike WFS mask.

Following the ZWFS mask is a lens of focal length 30 mm that collimates the beam. However, for faint objects an option will be provided to substitute a 15mm focal length lens, which will give a narrower collimated beam and more compact images on the detector, so minimising detector noise.   The collimated beam is then directed to the down periscope by a knife-edge mirror. Given that star images are to be formed at the cold stop and pupil images on the sensor, a re-imaging lens is necessary. Moreover, with the four beams all forming star images at the same place (cold stop) and with the input beams being collimated, it follows that the beams coming from the knife-edges to the re-imaging lens must be parallel to each other. To obtain adequate separation of the beams so that one knife-edge does not vignette another’s beam, and to obtain good separations of the spots on the sensor, the upper periscope mirror and re-imager must use 2 inch optics (the chief rays of the outer beams of the four are $\pm 11.54$ mm off-axis). The design shown here satisfies the layout requirements and leaves space for Bifrost optics and the Solarstein calibration unit, while directing the output beams to the same C-RED One as used for Heimdallr. The re-imaging lens in the down periscope is a custom doublet, optimised for operation at J band as well as H, as required for the optional use of Bifrost in H band (when Baldr would use J band).

\begin{figure}[htbp]
\centering\includegraphics[width=0.8\textwidth]{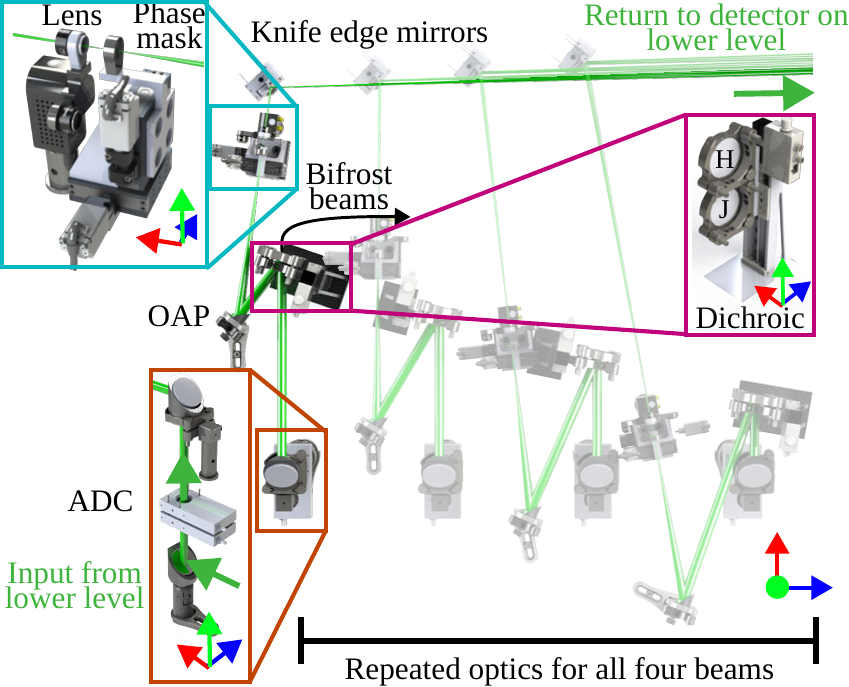}
\caption{Top view of the Baldr optical layout used to sense the wavefront of each beam. The beams are incident on the dichroic from the lower level (orange inset) and an \gls{ADC} is mounted on the underside of the breadboard (omitted for clarity). The dichroic transmits Bifrost beams to fold mirrors (not shown), and reflects either H or J band light to the OAP, depending on motor configuration. The OAP brings beams to focus on a phase mask that can be positioned in the plane perpendicular to the beam (blue inset). A flip mount enables changing between low and high \gls{SNR} configurations. A series of knife-edge mirrors positions the beams in a horizontal pattern and returns them to the Heimdallr detector on the lower level. }
\label{fig:baldr}
\end{figure}

\begin{figure}[htbp]
\centering\includegraphics[width=0.5\textwidth]{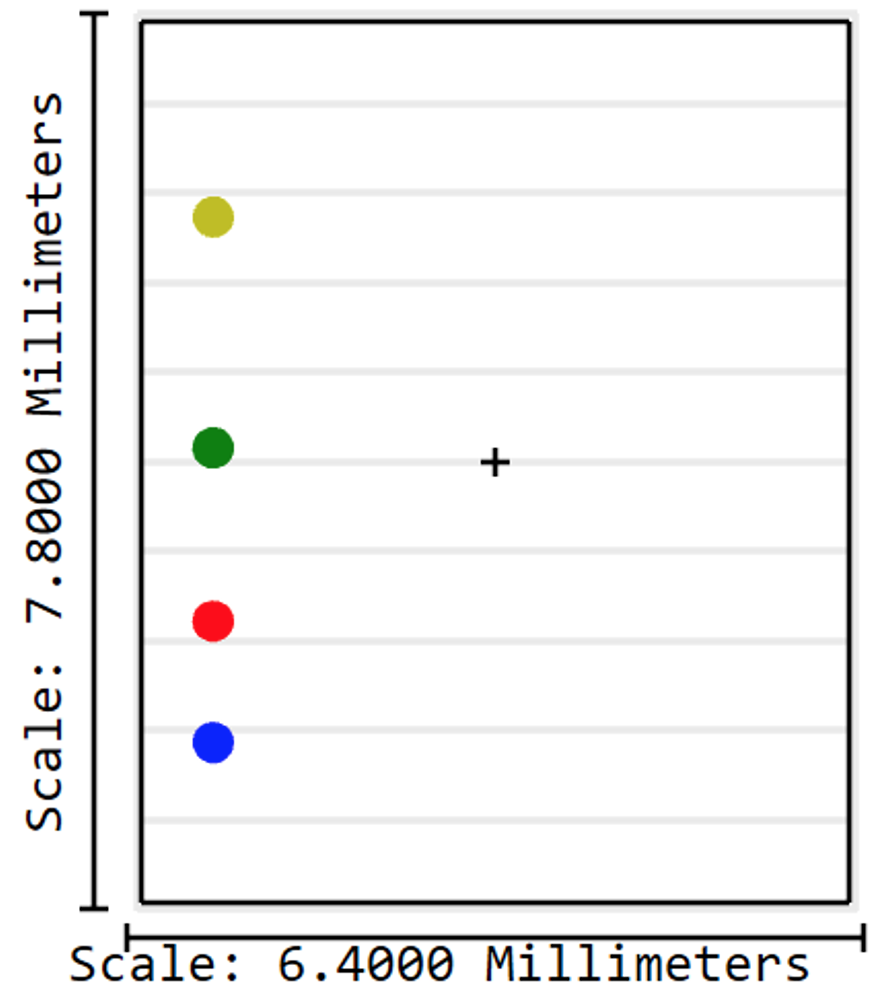}
\caption{Footprint diagram of the four Baldr pupil images on the sensor plane.}
\label{fig:Baldr_footprint}
\end{figure}

\autoref{fig:Baldr_footprint} shows the location and size of the four pupil images on the sensor plane. The spots have a diameter of $\sim 290 \mu $m. Their spacing is unequal because the knife-edge mirrors are arranged to provide greater spacing for configurations 3 and 4, to allow for diffraction broadening of the beams as they proceed towards the down periscope (see \autoref{fig:baldr}). This minimises the possibility of vignetting at the knife-edges.

\subsection{Baldr image quality}
\begin{figure}[htbp]
\centering\includegraphics[width=0.85\textwidth]{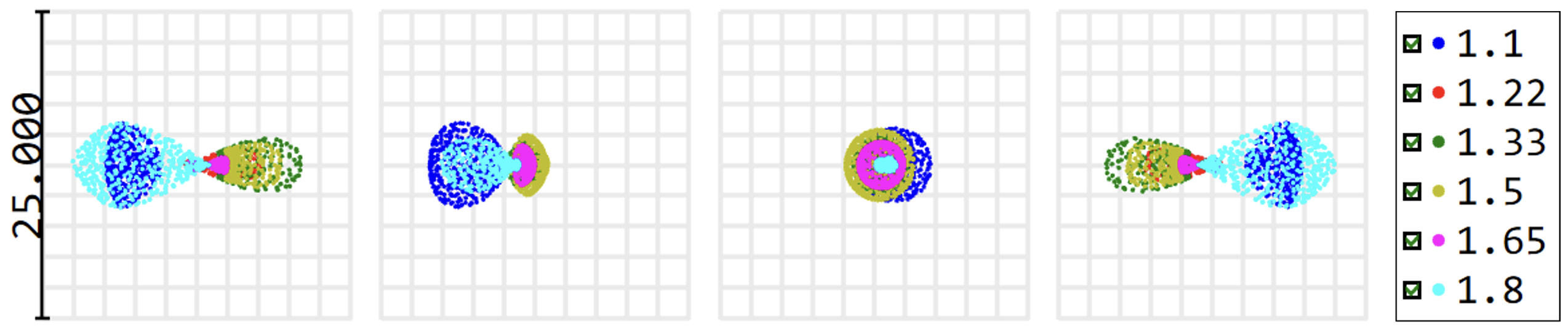}
\caption{Spot diagrams showing {\it pupil} image quality at the sensor plane. The box size is 25 $\mu$m on a side; the Airy disc (for pupil imaging) is not shown because it is much larger than the box. The numerical values in the legend are wavelengths in $\mu$m.}
\label{fig:Baldr_IQ}
\end{figure}

Baldr forms pupil images on the sensor, not star images. 
In order to assess the accuracy of the pupil imaging a ray-trace simulation was carried out using the re-imaged pupil as a source. The results are shown in \autoref{fig:Baldr_IQ}. Wavefront errors are mostly less than 0.08 $\lambda$ but have maxima of 0.14 $\lambda$.

\subsection{Optomechanical design}

\autoref{fig:baldr} shows the optomechanical layout of Baldr. The positioning of the \gls{ADC} on the underside of the breadboard has two advantages. First, this setup is extremely compact, adding functionality in the (otherwise optically unused) upwards beam. Also, it can be installed in a later phase of development, along with the off-axis arm of Bifrost which requires the correction. 

To act as a reliable phase reference, the phase mask must be actuated in the directions perpendicular to the axis of the beam. A compact, inexpensive design using two Zaber linear actuators mounted on 1/2" translation stages at right angles was selected. This provides a repeatability of $1.5$ $\mu$m, which is much less than the $\sim40 \mu$m spot size in H band, enabling the critical requirement in the \gls{WFS} of placing the mask in the beam. In general, these observation mode changes are made by stepper motors. To meet power budget requirements, these are only moved in full steps and are powered off when not in use.

Immediately following the phase mask is the re-imaging lens. To change between points on the \gls{SNR} vs number of corrected modes trade-off, there is a choice of two re-imaging lenses. The flipper motor mount with the 1/2'' lens mount has a radius of curvature of $36.1$ mm and an angular repeatability of 50 $\mu$rad, and hence the lenses have a positional repeatability of 2 $\mu$m, which is sufficiently accurate to avoid off-axis aberrations. Lenses are threaded on opposite sides of the mount for a compact design. Shims are used to ensure the correct relative separation between the lenses. 

Finally, for the upper level as a whole, several important design decisions were made. The VLTI lab limits instruments to a total weight of 500 kg, which restricts the possible upper level designs. Two options were considered: using an UltraLight (aluminium) breadboard from Thorlabs, or using a breadboard with thinner steel sheets, such as that from Edmund Optics. Since the optical bench below is steel, differential thermal expansion between seasons (up to 15$^\circ$C difference in the lab) was considered in the aluminium breadboard design. Over distances on the order of a meter, the differential expansion is $\sim100$ $\mu$m. This could be acceptable under the assumption that the breadboard is able to expand linearly rather than bend. This could be achieved with an aluminium breadboard design employing breadboard seats featuring rollers constrained in only one dimension. Either design is valid but we selected the steel breadboard due to availability and lead times. 

We also consider the placement of the posts in order to spread the weight evenly around the centre of mass and avoid vignetting of beams on the lower level. Finally, the earthquake survivability of this design was verified. By considering the peak horizontal acceleration in the requirements of 0.34 g acting on the 85 kg upper level, the total force experienced is well within the strength of 16 M6 screws. 

\subsection{Summary}

Baldr is a Zernike \gls{WFS} well poised to act as a secondary \gls{AO} system for Asgard. Our design meets requirements including observing modes in the H or J bands through careful consideration of optical elements and motorized components. The result uses a majority of off-the-shelf components that are readily accessible, complemented with custom optics where needed. Ultimately, the design of Baldr presented here enables critical, reliable high order wavefront correction for a wide range of targets.

\section{Solarstein}
\subsection{Goals and requirements}

Solarstein is designed to enable the alignment and calibration of the suite, presenting as a telescope simulator for different stages of the commissioning and observing schedules. This module will interface with Asgard in the same way as the \gls{VLTI}, with four cophased, uniformly illuminated (within 10\%) beams in the wide waveband 1-4 $\mu$m, as well as a visible laser for alignment. The beams are high fidelity in terms of pupil location, secondary obstruction features and diameter for an unresolved source. Meeting these requirements enables Asgard to be aligned during commissioning and regularly calibrated during observing. The VLTI source MARCEL will be used to initially align Asgard, but it does not have bright enough continuum sources for J and H bands, and also does not include a wavelength calibration source. Additionally, Baldr has a goal of rapid re-calibration during night-time observations on moving the ADCs.

\subsection{Optomechanical design}

\begin{figure}[htbp]
\centering\includegraphics[width=0.85\textwidth]{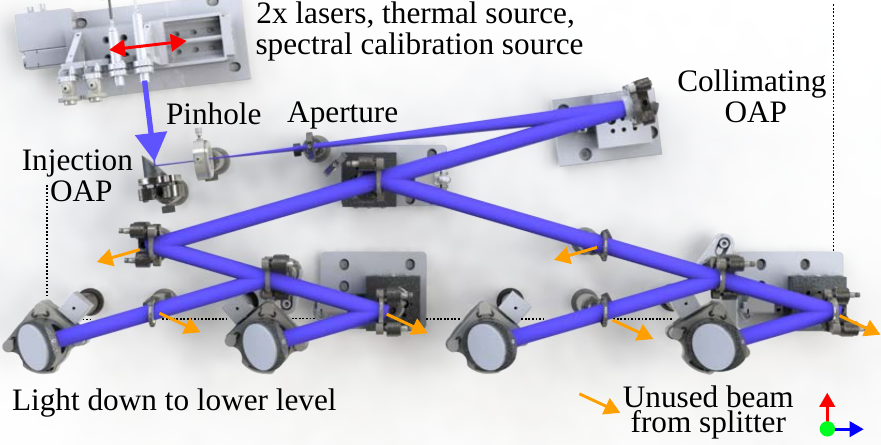}
\caption{Solarstein optical layout, with black dashed line indicating the edge of the upper level breadboard. Using the stage at the top left, a source is selected. The light is injected with an \gls{OAP} through a pinhole, creating an unresolved source. The diverging beam has an aperture stop with a secondary obstruction. Another \gls{OAP} then collimates the beam at 18mm diameter. Now the beam is split into four identical, cophased beams using a series of beamsplitters, where each final beam is produced by two reflections and two transmissions. Each beam is then reflected down to the lower level with right-angle mirrors.}
\label{fig:solarstein}
\end{figure}

\autoref{fig:solarstein} shows the layout of Solarstein on the upper level of the Asgard suite. There are five main components: source stage, pinhole injection, collimation, splitting and the interface to lower level. 

First, Solarstein has four sources: a tungsten lamp for \gls{IR} use, a spectral calibration source for Bifrost and two lasers for alignment. Two visible lasers are used with different wavelengths to ensure some transmission occurs through \gls{IR} dichroics, which typically have poorly defined behaviour at these wavelengths. 

Next, a pinhole is used to create an unresolved source. The $f$/25 beam exiting from the pinhole means that to be unresolved at 1\,$\mu$m wavelength, the pinhole should be smaller than 25\,$\mu$m - a 20\,$\mu$m pinhole was chosen. Note that a partially resolved pinhole does not reduce fringe visibility as long as downstream filters on each beam are aligned to the same part of the pinhole image. To collimate the resulting beam, a custom \gls{OAP} is used, which, for manufacturing, matches OAP2 in Heimdallr. The \gls{VLTI} pupil is 2510 mm from the edge of the AMBER table, so (after taking into account the path length through the rest of Solarstein) the secondary obstruction must be 333 mm from the collimator \gls{OAP}. This places it within a lens tube distance of the pinhole, and hence only a single mount is used.

After the beam is collimated, a series of beamsplitters produce four copies of the beam, using the paths that have two reflections and two transmissions. The substrate used is a 30 arcmin wedged CaF$_2$ round with a 6 nm coating of nickel. A wedge ensures ghost reflections are not present. To avoid chromatic shifts, the beamsplitters are clocked and oriented in a way such that the wedges of the desired beams have destructive combinations. This array has two degrees of freedom without changing any relevant optical properties: the angle of incidence on each beamsplitter and the distance from the collimator OAP to the first beamsplitter. These were varied using equations in Solidworks and placed to minimise table space whilst avoiding vignetting and ensuring there is no clash with Baldr. Three of the beamsplitters are placed on translation mounts with linear actuators, used for fringe search in the cophasing stage. 

Finally, a right-angle mirror mount directs the light downwards to the lower level, with a flip mirror controlling if the light is injected to Heimdallr. The right-angle mirrors are cantilevered over the edge of the breadboard since periscope holes cannot be drilled within 50mm of the edge, and the perspex enclosure around the instrument must rest on the bench as a whole. 

An alternative design we considered involved sampling sub-apertures of a larger, collimated beam as is done in the six telescope simulator \cite{sts}. Such a design is achromatic, however, due to the profile of fibres feeding the collimated beam, it has a non-symmetric, non-uniform illumination over the sub-apertures. The beamsplitter design presented in this work has a lower, wavelength-dependent throughput, but contains degrees of freedom to tradeoff uniformity of the illumination pattern (which is always symmetric when aligned) with throughput. We note that the beamsplitter architecture is much more compact and permits easier cophasing between beams by back collimation. Thus, this beamsplitter design is selected over the sub-apertures design.

\subsection{Summary}

Solarstein serves as the alignment and calibration unit for the entire Asgard suite, presenting a set of four cophased, uniformly illuminated beams that simulate the VLTI. For alignment, standard interferometric use and spectroscopy Solarstein contains laser, thermal and spectral calibration sources respectively. We leverage a novel beamsplitter array to keep the overall design more compact than the alternative (sub-apertures from a larger collimated beam). The design also provides a direct way to verify the beams are cophased through back collimation.  

\section{Discussion}\label{sec:discussion}
\subsection{System-wide insights} \label{sec:system_insights}
In formulating the architecture for this large system, we have developed a new workflow that saves time and improves quality checks on designs as they progress. In particular, we found cases early on in the design where feedback was needed on both the optical quality (from Zemax) and the clearances to mount edges (from Solidworks). Typically, a designer exports Zemax to a \gls{CAD} file (using the built-in STEP export) and positions mounts with respect to these parts using mates with exported surfaces. The mount to optic relation, however, doesn't persist through updates to the Zemax design since the STEP is completely new, resulting in the need to manually redifine mates at every design iteration. We developed a code that scans the prescription data in Zemax for the surfaces and exports relevant positions and orientations to a text file to be read by Solidworks as equations. The Solidworks mates, now linked to the text file, persist even when the STEP files are updated. Furthermore, this framework enables simple transformations on a per-surface or global coordinate basis, giving greater flexibility for experimentation. We believe this workflow is particularly effective for rapid prototyping with off-the-shelf parts, where opto-mechanical constraints are more acute.

In addition, Section 4 of the supplemental document describes methods used in the optical design to work with and align multiple configurations, as well as a sketch of the calculations to steer and place knife-edge mirrors.

\subsection{System integration}
We now turn to a sketch of the way in which such a system could be aligned and referenced for testing. 
In the use cases not at the \gls{VLTI}, such as the integration testing in Nice, France planned for the middle of 2024, Solarstein is first aligned using the laser to produce four beams on the lower level in the correct position and height. Next, we change to the thermal source, back collimate the light using mirrors mounted on the lower level table holes as a fiducial and perform fringe search until beams are cophased. From here, remaining instruments can be aligned sequentially, starting with Heimdallr, using the laser sources. Integration testing on the assembled system can then be conducted using the thermal source.
When at the \gls{VLTI}, Solarstein must be a direct substitute for the existing low throughput source, MARCEL. Hence, rather than cophasing using back collimation, only laser alignment of Solarstein and Heimdallr is performed. After alignment, Heimdallr is cophased to MARCEL. Next, MARCEL is swapped for Solarstein and beams are cophased while holding Heimdallr fixed. The remainder of the procedure continues as in the previous case. 

For this instrument suite the commissioning plans and corresponding observing runs are split into phases. In phase 1, Heimdallr, Baldr, Nott and one injection module of Bifrost will be assembled at \gls{VLTI}, with the instruments operating in a subset of the observing modes. In phase 2, the full functionality of Heimdallr and Baldr on the AT's will be demonstrated. Finally, in phase 3, the full instrument suite will be integrated and have the capabilities to work on the UT's.

\section{Conclusion}
In this work, we have presented Heimdallr, Baldr and Solarstein -- three critical modules at the heart of the next-generation Asgard suite recently approved by \gls{ESO}. Heimdallr, the instrument responsible for beam combination in K band, contains common optics with internal delay lines and images a non-redundant pupil in two wavebands to infer relative phases and low order aberrations from the telescope beams. A novel design for the rejection of thermal noise is presented in the form of a ``Narcissus box'' containing optics which reflect the cold stop onto itself at all unused pupil positions and wavelengths. Baldr, a 4-beam adaptive Zernike wavefront sensor system uses the same detector and fast actuators as Heimdallr with the deformable mirror acting as a piston and adaptive optics actuator. Internal calibration and alignment is enabled by the Solarstein source. Although complex, the modularity and benchtop nature of these components of Asgard will enable comprehensive testing of the system both in Sydney and Nice prior to integration at ESO's Paranal observatory.

Taken together, these modules enhance the fringe tracking and adaptive optics speed of VLTI by a factor of $\sim$4, which is critical to enabling both short-wavelength interferometry with Bifrost and high contrast interferometry with Nott. Additionally, Heimdallr will be a powerful visibility and closure phase beam combiner optimised for K-band, initially available for community use through collaboration with the instrument team.
\begin{backmatter}
\bmsection{Funding}

This research was supported by the Australian Research Council (ARC) Linkage Infrastructure Funding (LIEF) grant LE220100126.

\bmsection{Acknowledgements}
We would like to thank Barnaby Norris, Connor Langford and Lucinda Lilley for their helpful suggestions regarding figures. 

This work used the ACT node of the NCRIS-enabled Australian National Fabrication Facility (ANFF-ACT).

\bmsection{Disclosures}
\noindent The authors declare no conflicts of interest.

\bmsection{Data Availability Statement}
\noindent No data were generated or analyzed in the presented research.
See Supplement 1 for supporting content.

\end{backmatter}

\bibliography{main}

\end{document}


\maketitle

\section{Use of off-axis paraboloids in Zemax}

\begin{figure}[htbp]
\centering\includegraphics[width=0.8\textwidth]{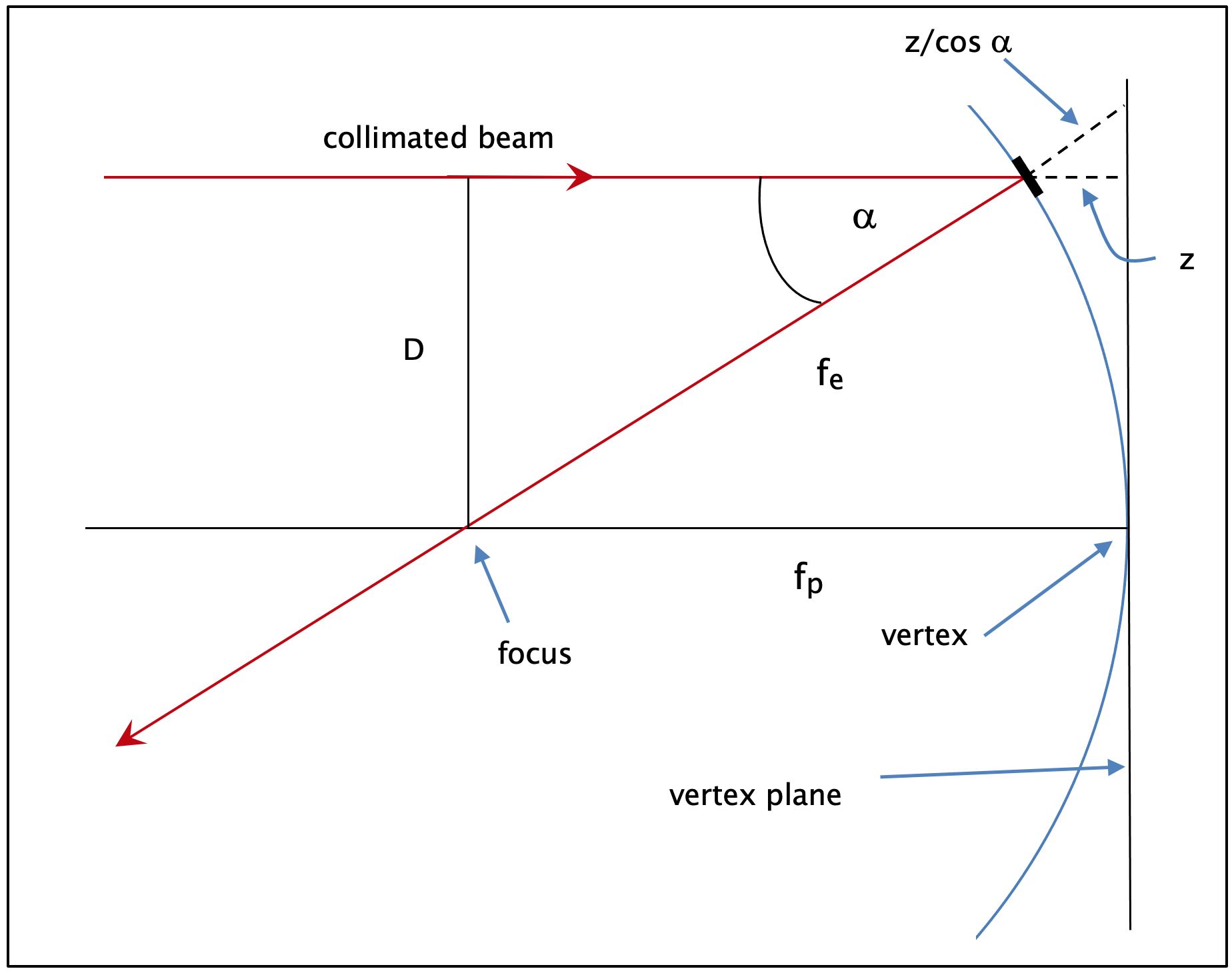}
\caption{Off-axis paraboloid geometry, with the full paraboloid shown in blue.}
\label{fig:OAP_geom}
\end{figure}

Figure \ref{fig:OAP_geom} shows a collimated beam encountering a paraboloid, parallel to the axis. Only the part of the paraboloid near the point of incidence will actually be built, {\it i.e.} the small segment will be an off-axis paraboloid (OAP).
There are two fundamental parameters which describe the inherent properties of an OAP:
\begin{itemize}
    
\item f$_{\rm p}$ is the focal length of the parent sphere. It is half (the modulus of) the radius of curvature of the sphere (‘radius’ in Zemax).
\item
Either $\alpha$ or D. $\alpha$ is the off-axis angle, while D is the offset of the OAP from the paraboloid’s axis. They are fixed once the OAP has been manufactured, because the OAP must always be used with the collimated beam parallel to the axis – otherwise aberrations would be severe. D will be used in Zemax as a decenter, to get the paraboloid’s vertex appropriately offset from the collimated beam.
\end{itemize}
The derived parameters are:
\begin{itemize}
\item
f$_{\rm e}$ is the effective focal length of the OAP – {\it i.e.} the distance from the point of incidence of the collimated beam to the focus. If $\alpha$ is small then f$_{\rm e}$ is very close to f$_{\rm p}$ but it can be significantly more when $\alpha$ is larger. It is given by
\begin{equation}
    f_e = \frac{2f_p}{\sin^2 \alpha}(1 -\cos \alpha)
\end{equation}
\item $D$ and $\alpha$ are related by
\begin{equation}
    D = f_e \sin \alpha
\end{equation}
or
\begin{equation}
    \tan \alpha = \frac{D}{\left(f_p - \frac{D^2}{4f_p} \right)} =  \frac{D}{(f_p - z)}
\end{equation}
where $z$ is defined below.
\end{itemize}

In Zemax, the ‘thickness’ values both before and after the OAP are distances to the vertex plane, not to the OAP itself. 
The collimated beam encounters the OAP surface at a distance from the vertex plane equal to the ‘sag’ which is 
\begin{equation} 
z = \frac{D^2}{4f_p}
\label{eqn:sag}
\end{equation}

The beam from the OAP to the focus has Zemax thickness 
\begin{equation}
    {\rm thickness} = f_e + z/\cos \alpha
    \label{eqn:thickness}
\end{equation}
However, note that making correct use of the  above numerical values requires attention to the local coordinate system. Thickness is measured along the local $z$ axis, so it is necessary to ensure that this is along the chief (central) ray of the beam. This can be done by using a coordinate break after the OAP, with ‘chief ray’ solves for the necessary decentres and tilts.

Equation (\ref{eqn:thickness}) is clearly not applicable when $\alpha$ = 90$^\circ$, which is quite possible with off-the-shelf 90$^\circ$ OAPs. In that case, the thickness for the collimated beam can still be calculated with the extra sag from equation (\ref{eqn:sag}), but the distance from the OAP surface to the focus is not given by equation (\ref{eqn:thickness}). In this case the separation arises from the size of the optic, which is tilted by 90$^\circ$ and decentred by an amount equal to f$_e$.

\section{Heimdallr components}
The Heimdallr components are listed in Table \ref{fig:Table_1_1}
\begin{table}[htbp]
\centering\includegraphics[width=1.0\textwidth]{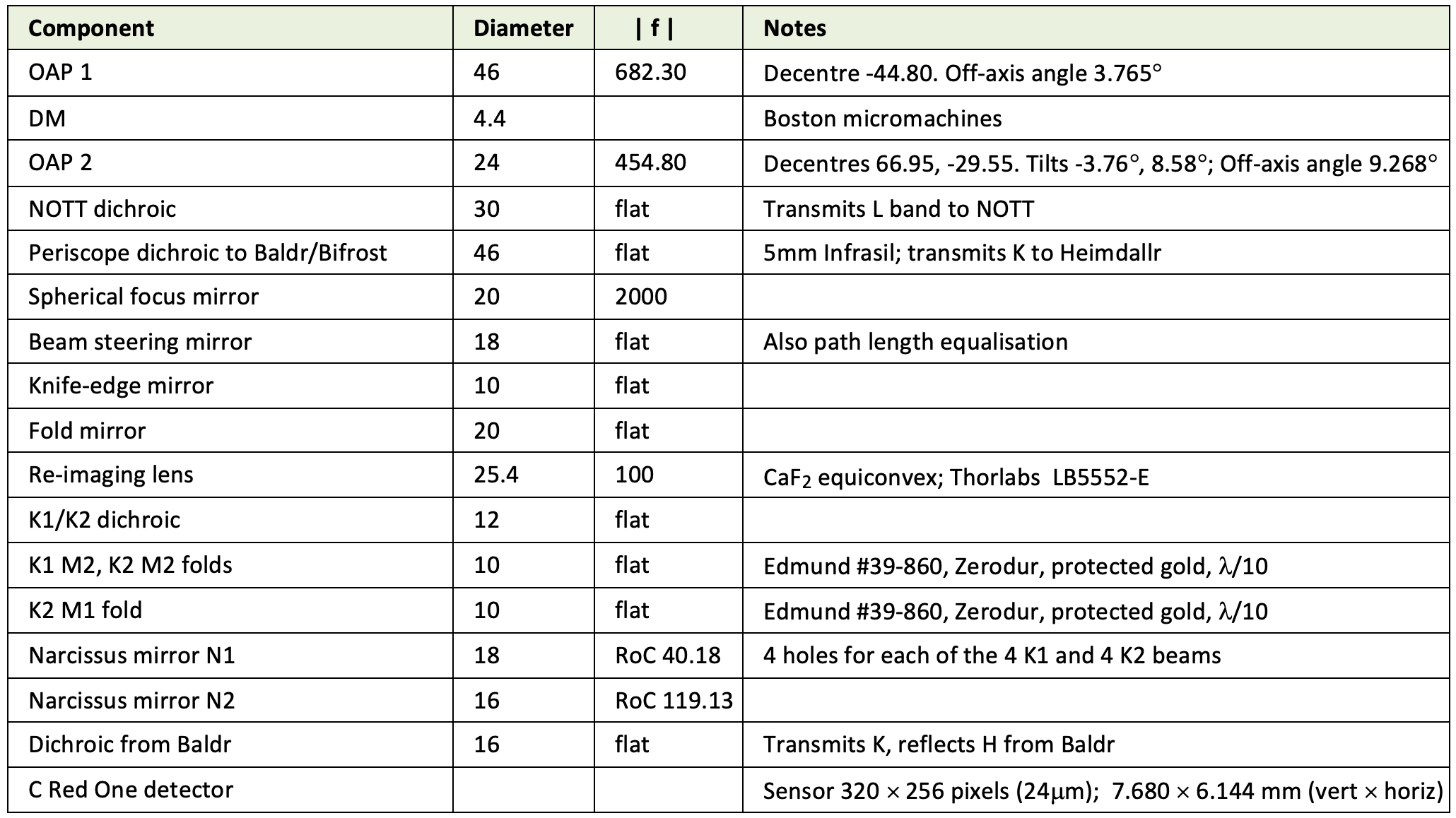}
\caption{Components of the Heimdallr optical train. All dimensions are in mm. Column 3 gives the focal length of powered optics. }
\label{fig:Table_1_1}
\end{table}

\vspace*{2cm}
\section{Baldr components}
The Baldr components are listed in Table \ref{fig:Baldr_Table_9_1}
\begin{table}[htbp]
\centering\includegraphics[width=1.0\textwidth]{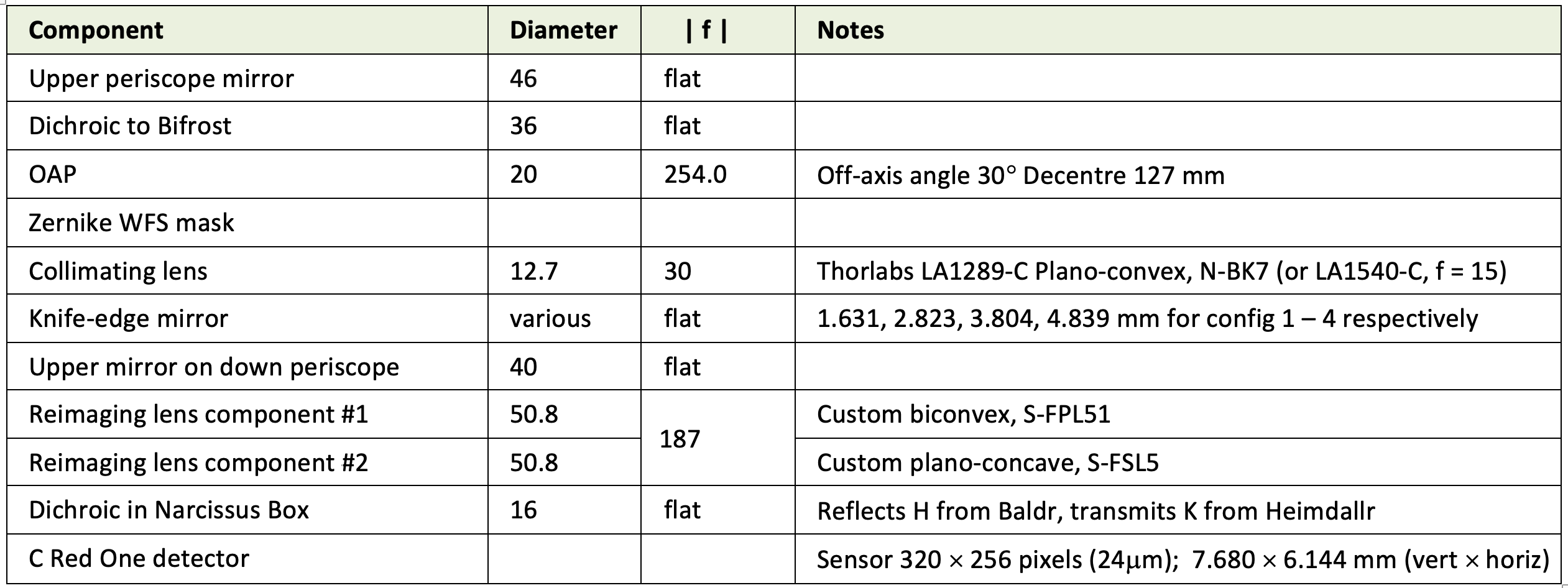}
\caption{Components of the Baldr optical train. All dimensions are in mm. Column 3 gives the focal length of powered optics. }
\label{fig:Baldr_Table_9_1}
\end{table}

\section{Optical design in Zemax}
This section describes some methods used in the optical design and its translation to Zemax.

\subsection{Configuration offsets}
The four configurations are offset from each other using decenters that precede a surface designated as the global coordinate reference, then the decenters are reversed after the reference surface.  

\subsection{Global vertex data}
Each optical element is placed by Zemax in accordance with the local coordinate system, which changes with each decenter or tilt. But for consistency rays are traced in the single global coordinate system, and this is what is shown in the 3D Layout diagrams. A very useful facility is the Global Vertex listing within the Prescription Data. This lists the XYZ positions of the vertex of each surface, and the XYZ tilt angles, in the fixed global coordinate system.

Note that for an OAP, the Global Vertex gives the position of the parent surface's vertex, not the offset OAP itself.

Since Baldr's beams have to rejoin the Heimdallr system at the Narcissus Box, it was convenient to use the same global coordinate reference for Baldr and Heimdallr.

\subsection{Alignment of multiple configurations}
Both Heimdallr and Baldr begin with four separate configurations, with each having its own mirrors, lenses etc. Then the beams come together after the knife-edge mirrors, and all encounter the same single optical elements ({\it e.g.} the Heimdallr fold mirror and re-imaging lens). But in Zemax these are still separate configurations, each with its own mirror and lens. It is thus necessary to align these to sufficient accuracy so the 8 configurations in Heimdallr (4 telescopes $\times$ 2 for K1/K2) can act as a single optic. This is done using tilts and decenters at a conveniently placed alignment surface, and using data from the `Global Vertex' listings. The aim is to make all the local coordinate systems coincident in both position and tilts, having selected one of the beams as the fiducial.

At the alignment surface, the decenters and tilts are naturally in the local coordinate system, while the alignment data from Global Vertex are in global coordinates. Thus it will generally be necessary to apply coordinate system shift and rotation to calculate the required corrections at the alignment surface.

\subsection{Placing and aiming the steering and knife-edge mirrors}
The Heimdallr knife-edge mirrors are not in a single plane, since they have to create the 2D triad of beams around the central fiducial beam. The knife-edges are placed by suitable tilts and thicknesses at the preceding steering mirrors, which were calculated in an Excel spreadsheet. Each knife-edge was constrained to a line on the surface of a cone with opening angle 0.758$^\circ$ (1.04$^\circ$ after the re-imaging lens), then Excel goal-seek was used to place it along this line at the position that gives path length equality with the fiducial beam.

The steering mirrors and knife-edges all needed suitable tilts in order to aim their beams to the desired locations. Since tilts are applied in local coordinates, while aiming error data were in global coordinates, a conversion was necessary. This was done by applying small test tilts, separately about X and Y, and noting their effects at the subsequent surface. This provided the required partial derivatives, which were then formulated into a matrix equation giving the effect in global coordinates ({\it e.g.} at the knife-edge position from X and Y tilts at the steering mirror). A matrix inversion solution then produced the required X and Y tilts to achieve the desired global position of the knife-edge.
